\documentclass[AMA,LATO1COL]{WileyNJD-v2}

\usepackage{amsmath,amssymb,amsfonts}
\usepackage{graphicx}
\usepackage{textcomp}
\usepackage{xcolor}
\usepackage[utf8]{inputenc}
\usepackage{booktabs}
\usepackage{array}
\usepackage[]{hyperref}
\usepackage[rightcaption]{sidecap}

\articletype{Special Issue Paper}
\received{--}
\revised{--}
\accepted{--}

\begin{document}

\title{%
    Real-Time XFEL Data Analysis at SLAC and NERSC: a Trial Run of Nascent
    Exascale Experimental Data Analysis%
}

\author[1]{Johannes P. Blaschke*}
\author[2]{Aaron S. Brewster}
\author[2]{Daniel W. Paley}
\author[2]{Derek Mendez}
\author[2]{Asmit Bhowmick}
\author[2]{Nicholas K. Sauter}
\author[3]{Wilko Kröger}
\author[3]{Murali Shankar}
\author[1]{Bjoern Enders} 
\author[1]{Deborah Bard}

\authormark{BLASCHKE \textsc{et al}}

\address[1]{%
    \orgdiv{National Energy Research Scientific Computing Center},
    \orgname{Lawrence Berkeley National Laboratory},
    \orgaddress{1 Cyclotron Road, Berkeley, CA 94720},
    \country{USA}%
}

\address[2]{%
    \orgdiv{Molecular Biophysics and Integrated Bioimaging Division},
    \orgname{Lawrence Berkeley National Laboratory},
    \orgaddress{1 Cyclotron Road, Berkeley, CA 94720},
    \country{USA}%
}

\address[3]{%
    \orgname{SLAC National Accelerator Laboratory},
    \orgaddress{2575 Sand Hill Road, Menlo Park, CA 94025},
    \country{USA}%
}

\corres{%
    *Johannes P. Blaschke,
    Lawrence Berkeley National Laboratory,
    1 Cyclotron Road, Berkeley, CA 94720, USA.
    \email{jpblaschke@lbl.gov}%
}

\abstract[Summary]{X-ray scattering experiments using Free Electron Lasers
    (XFELs) are a powerful tool to determine the molecular structure and
    function of unknown samples (such as COVID-19 viral proteins). XFEL
    experiments are a challenge to computing in two ways: i) due to the high
    cost of running XFELs, a fast turnaround time from data acquisition to data
    analysis is essential to make informed decisions on experimental protocols;
    ii) data-collection rates are growing exponentially, requiring new scalable
    algorithms. Here we report our experiences analyzing data from two
    experiments at the Linac Coherent Light Source (LCLS) during September
    2020. Raw data were analyzed on NERSC’s Cori XC40 system, using the
    Superfacility paradigm: our workflow automatically moves raw data between
    LCLS and NERSC, where it is analyzed using the software package CCTBX\@. We
    achieved real time data analysis with a turnaround time from data
    acquisition to full molecular reconstruction in as little as 10
    min~--~sufficient time for the experiment’s operators to make informed
    decisions.  By hosting the data analysis on Cori, and by automating
    LCLS-NERSC interoperability, we achieved a data analysis rate which matches
    the data acquisition rate. Completing data analysis within 10 mins is a
    first for XFEL experiments and an important milestone if we are to keep up
    with data-collection trends.
}

\keywords{%
    Distributed, Parallel, and Cluster Computing;
    Real-time, and Urgent High-Performance Computing%
}

\maketitle

\section{Introduction}

\begin{figure*}
\centering{}
\includegraphics[width=1.00\textwidth]{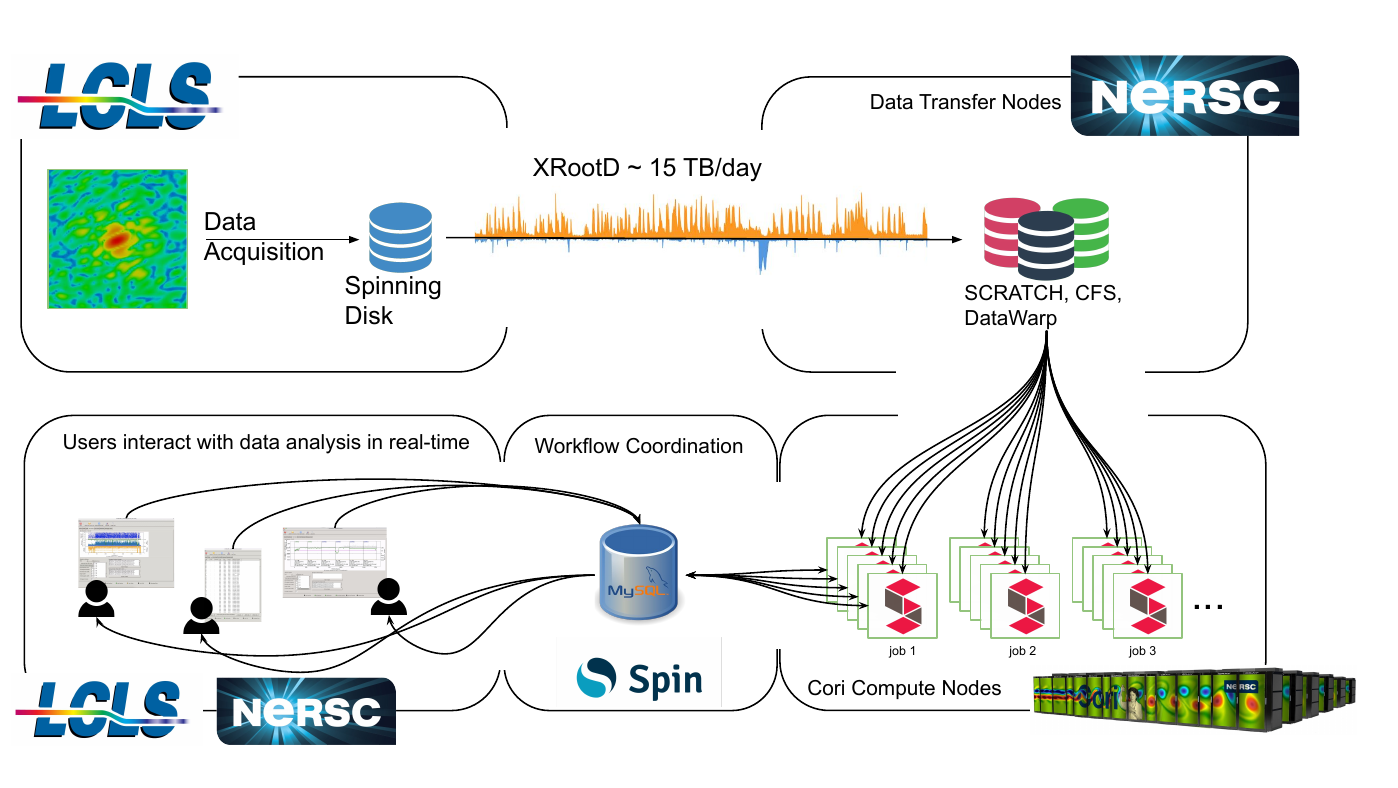}
\caption{%
    Sketch of the Superfacility workflow: \emph{Top:} Data are automatically
    transferred from the LCLS spinning-disk storage system via XRootD to
    NERSC's Scratch file system (the orange and blue spikes show the data
    transfer rate into and out of NERSC, respectively~--~spike height ranging
    from approx. 1.3 to 2.6 GB/s~--~over the ESNet network during the same time
    as the experiment, with each spike being a completed run). \emph{Bottom}:
    At NERSC the CCTBX workers (running in Shifter containers on the Cori
    compute nodes) automatically analyze new data on Scratch, using the
    DataWarp burst buffer as a cache. Users at LCLS and NERSC connect to a
    MySQL database hosted at NERSC to orchestrate the workers, review the data
    analysis and iterate analysis parameters.%
    \label{fig:workflow}%
}
\end{figure*}

X-ray scattering experiments using Free Electron Lasers (XFELs) are a powerful
tool to determine the molecular structure and function of unknown samples, such
as COVID-19 viral proteins. The X-ray light produced by XFELs is particularly
useful as a tool for probing microscopic samples as it is coherent and intense,
allowing teams of scientists to probe structural details that leave only a weak
trace signal \cite{sauter2020}. However all of this comes at a significant
cost: XFEL facilities require specialized equipment and large teams to operate.
To operate efficiently, it is essential that the experimental investigators
have immediate feedback from data analysis in order to make informed decisions
about their experiments in real time. By 2025 the next generation of XFEL
experiments will more than double the detector resolution, and increase the
rate at which measurements are taken by a factor of over 400$\times$ compared
to existing facilities\cite{superfacility,bard2022}. This will require
computational intensity levels to escalate from petascale to exascale, for data
analysis to keep pace with data collection.

To rise to these challenges, the Linac Coherent Light Source (LCLS) at SLAC has
partnered with the National Energy Scientific Computing center (NERSC) at LBNL
using a ``Superfacility'' model \cite{superfacility,blaschke2023}: data
collected at SLAC are immediately transferred to NERSC (via ESnet) where they
are analysed on the Cori XC40 supercomputer%
\footnote{%
  \url{https://docs.nersc.gov/systems/cori/}%
}.%
The results are then reported back to the experiment's operators in real time.
In this paper, we demonstrate the usefulness of this approach by reporting our
experiences from two experiments in September 2020: LV95, which consisted of
small molecules related to materials science\cite{schrieber2022}; and P175,
which consisted of COVID-19 viral proteins and potential bound ligands
\cite{Keable2021}.  These experiments needed to test many samples during
limited beam-time.  In order to know when to move on to the next sample and to
make changes to experimental protocol, a complete (or near complete) analysis
of the collected data needs to happen at the same rates at which the data are
collected.

\section{Analysing LCLS data at NERSC}

Data were collected at a peak rate of 120 images/second (approx. 1/42 of the
data-collection rate expected after future light source upgrades), totalling 15
TB/day.  A total of 130 TB of raw data comprising 28 million images were
collected during the experiments described in this paper. This is too much data
to manage manually, therefore we use the Superfacility paradigm: our workflow
automatically moves raw data between LCLS and NERSC, where it is analyzed using
the CCTBX software package%
\footnote{%
    The scripts to build CCTBX at NERSC, and the Docker image used for the data
    processing jobs are available here:
    \url{https://gitlab.com/NERSC/lcls-software/-/tree/beamtime-2020-09/cctbx-production}%
    \label{ftn:docker_images}%
}%
\cite{grosse2002,sauter2013}.
By running on 64 Haswell nodes%
\footnote{%
    Each ``Haswell node'' is equipped with dual sockets. Each populated by an
    Intel Xeon 2.3 GHz 16-core E5-2698 v3 ``Haswell'' processor. Each node has
    128 GB DDR4 2133 MHz memory (four 16 GB DIMMs per socket). We note that
    even though NERSC's newest Supercomputer ``Perlmutter'' was not used during
    the beamtimes reported here, \emph{cctbx.xfel has been successfully
    deployed on Perlmutter\cite{blaschke2023}. The software deployment on
    Perlmutter has been successfully demonstrated during LCLS beamtimes.
    Furthermore, the docker images used here (\emph{cf.}
    \ref{ftn:docker_images}) are fully portable from Cori to Perlmutter's CPU
    nodes.} %
}%
, we achieved real time data analysis with a 10 min peak turnaround time from
data acquisition to full molecular reconstruction~--~sufficient time for the
experiment’s operators to make informed decisions between data-collecting runs.
At this computational intensity, the data analysis rate matches the data
acquisition rate. This demonstrates the usefulness of the Superfacility
approach: by automating job submission and data management, we were able to
analyze critical measurements within 10 mins, and most data in under 20 mins, a
first for XFEL experiments and an important milestone if we are to keep up with
instrument data-collection trends.

In this paper we give a detailed step-by-step description showing how our
workflow is deployed on NERSC's systems; how it coordinates data movement
(between SLAC and NERSC, discussed in section~\ref{sec:xrootd}) and data
analysis (via batch jobs at NERSC, discussed in
section~\ref{sec:data-analysis}); and how CCTBX enables interactive data
analysis with several human operators in the loop (discussed in
section~\ref{sec:pipeline-management}). CCTBX (specifically the
{\it{cctbx.xfel}}\ sub package) is a fully-automatic pipeline management system
that: i) tracks new incoming data, and relates it to experimental parameters
(``tags'') provided by the scientists; ii) automatically submits new analysis
jobs (using containerized workers) as new data come in; and iii) reports
analysis results via a database hosted on NERSC's ``Spin'' micro-services
platform in real time (discussed in section~\ref{sec:workflow-orchestration}).
This allows a team of scientists to work on the same data via one integrated
GUI, while CCTBX coordinates a ``swarm'' of workers behind the scenes.
Fig.~\ref{fig:workflow} illustrates this workflow.

\begin{table}
    \center
    \begin{tabular}{lllll}
    \toprule
        Experiment            & \multicolumn{2}{c}{LV95} & \multicolumn{2}{c}{P175}\\
        \midrule
        Spot finding          & 17M  & 6\%  & 11M  & 49\% \\
        Indexing              & 2M   & 25\% & 582K & 7\%  \\
        Refinement            & 564K & 99\% & 46K  & 85\% \\
        Integrating           & 559K & 99\% & 33K  & 97\% \\
        \midrule
        Total CPU utilization & \multicolumn{2}{c}{22663 core-hr} & \multicolumn{2}{c}{31167 core-hr} \\
        \bottomrule
    \end{tabular}
    \caption{%
        Sizes of the data sets collected during two experiments at the LCLS
        (LV95, and P175) as well as the size of different data analysis stages
        (described in section-\ref{sec:data-analysis}). The percentages show
        the average ``success rate'' for each stage~--~\emph{i.e.} the
        percentage of images to which the algorithm could find valid solutions
        (and thus can be used as inputs to the next stage). We use ``M'' to
        denote ``millions'' and ``K'' to denote ``thousands'' of diffraction
        images. Each image has a resolution of approx. 4 megapixels, requiring
        approx. 8 megabytes of storage. The LV95 data set is available at
        \url{https://dx.doi.org/10.11577/1839200}%
        \label{tab:processing}%
    }
\end{table}

\subsection{Transferring Data to NERSC}
\label{sec:xrootd}

\begin{SCfigure}[0.6]
    \centering
    \includegraphics[width=0.65\textwidth]{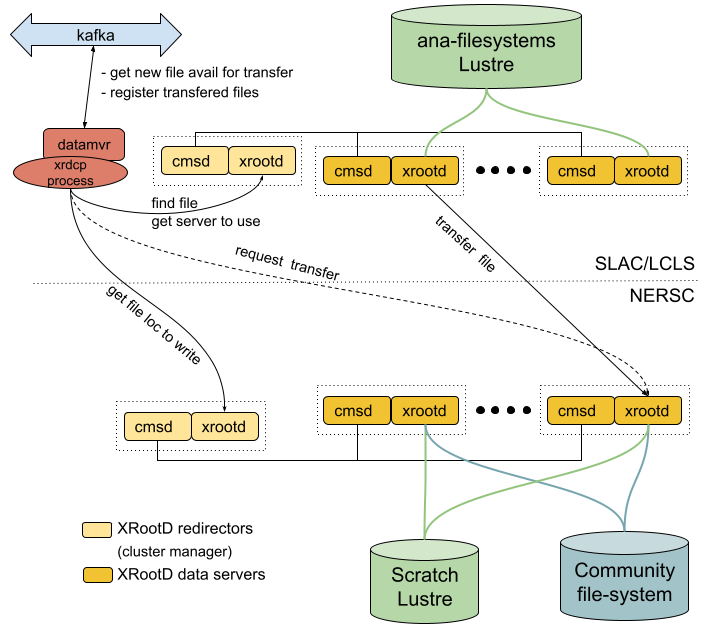}
    \caption{%
        Schematic of how the data mover transfers data using the NERSC~--~LCLS
        XRootD clusters. \emph{Top:} Kafka + data mover pipeline at LCLS
        together with the XRootD cluster used to send data (via ESNet) to the
        corresponding cluster at NERSC. \emph{Bottom:} XRootD cluster deployed
        on two data transfer nodes at NERSC\@. Once a new file is created, and
        logged as a file creation event in Kafka (the LCLS data ``logbook''
        service), the data mover initiates a data transfer using the XRootD
        cluster running at LCLS\@. The data is transferred via ESnet to its
        counterpart at NERSC, where the data is deposited in the \emph{SCRATCH}
        Lustre file system. Once a file has been transferred, its status in
        Kafka is recorded as ``available at NERSC''~--~allowing
        \emph{cctbx.xfel} to begin data analysis.%
        \label{fig:xrootsetup}%
    }
\end{SCfigure}

The LCLS data movers are responsible for the transfer of the data between the
different storage resources used by LCLS\@. Within the LCLS systems the data
are moved from the data acquisition storage to the high performance
fast-feedback storage and the large long term analysis storage.  The data
acquisition uses local SSD based storage on each of its nodes. The
fast-feedback storage is a shared 560TB nvme-SSD based file system using WekaFS
and the analysis storage is a 4PB spinning disk based Lustre file system.  The
movers also perform the data transfer to the remote HPC sites currently
supporting NERSC and the SLAC Shared Scientific Data Facility (SDF). At NERSC
the data mover copies data directly to the \emph{SCRATCH file system. Cori
scratch was a Lustre file system designed for high performance temporary
storage of large files. It had 30 PB of disk space, an aggregate bandwidth of
>700 GB/sec I/O bandwidth, and was made up of 10000+ disks and 248 I/O servers.
} The data mover is a component of the LCLS data management systems and
communicates with other components by publishing and subscribing to streams of
events using Kafka. The main events for the datamover are subscribing to
new-files-created events and publishing that files have been transferred to a
particular storage resource. For the remote transfers the XRootD data server is
used. Each remote site exports its shared file system through XRootD which runs
on multiple data transfer nodes that each site provides. All servers at a site
are clustered into a single system using XRootD's clustering functionality. The
data mover uses the XRootD transfer tool {\it xrdcp}\ in third party copy mode.
The data are directly transferred between an XRootD server at the source and
destination, without involving the node the mover is running on. In this
instance the destination pulls the data from the source.
Fig.~\ref{fig:xrootsetup} shows the NERSC and LCLS XRootD setup.  The main
entry point into each cluster is the redirector (aka cluster manager).  It
redirects the client to the data server that should be used for reading and
writing the data.

The data mover is a Python application whose main task is to perform many
transfers in parallel. It has two options to discover which files to transfer:
either monitor the experiment folder for new files or subscribe to a Kafka
stream (the LCLS data ``logbook'' service) which signals that new files have
been created. As new files are created the mover adds them to its internal
persistent queue. The files are sorted by run and the oldest runs are
transferred first. A run typically consists of 12-18 files. A third of these
files contain the detector data and are between a few to 100 GB in size. For
each of these files there are two index files that allow random access to the
detector data. The size of the index files is less than 1\% of the detector
data files. The observed peak ESNet transfer speeds showed bursts of 2.6 GB/s
whenever runs were completed. This measurement is the effective disk-to-disk
transfer rate, including the time it takes to read data from the Lustre file
system at LCLS, and the write incoming data to the Lustre at NERSC. The bursts
are due to the data only being transferred once the runs are ``concluded'' in
Kafka.

The Kafka + data mover + XRootD pipeline is fully automated and scalable. Once
an experimental run is concluded (a ``run'' is usually 5-30 mins worth of data
collection) the raw data files, as well as index files and calibration data,
are automatically recorded in Kafka as ``ready to be transferred'' and this
pipeline will begin the transfer to NERSC\@. The transfer is usually completed
within 3 minutes and the status is updated in Kafka as ``available at NERSC''.
The XRootD cluster is fully scalable, allowing us to transfer all the files
generated in one run at once.

\subsection{Pipeline Management}
\label{sec:pipeline-management}

\begin{figure*}
    \centering
    \includegraphics[width=0.494\textwidth]{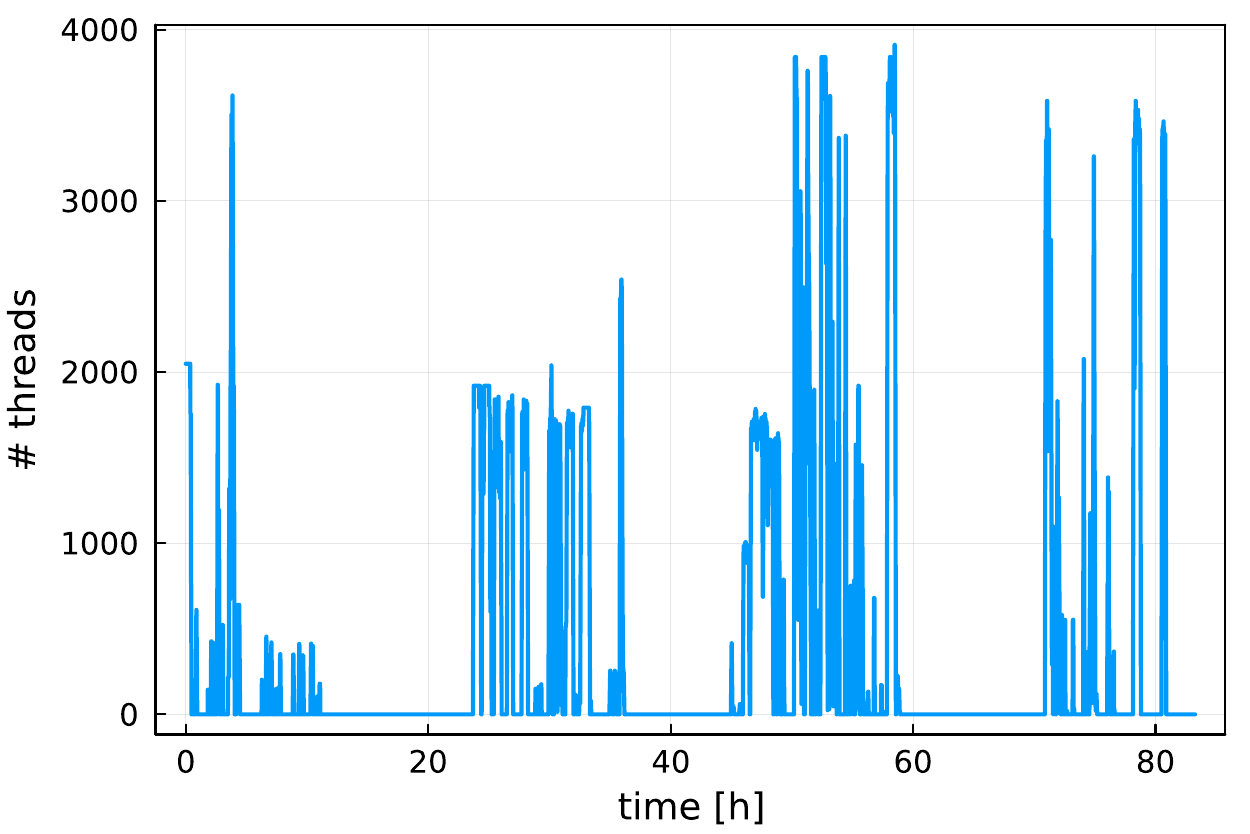}
    \includegraphics[width=0.494\textwidth]{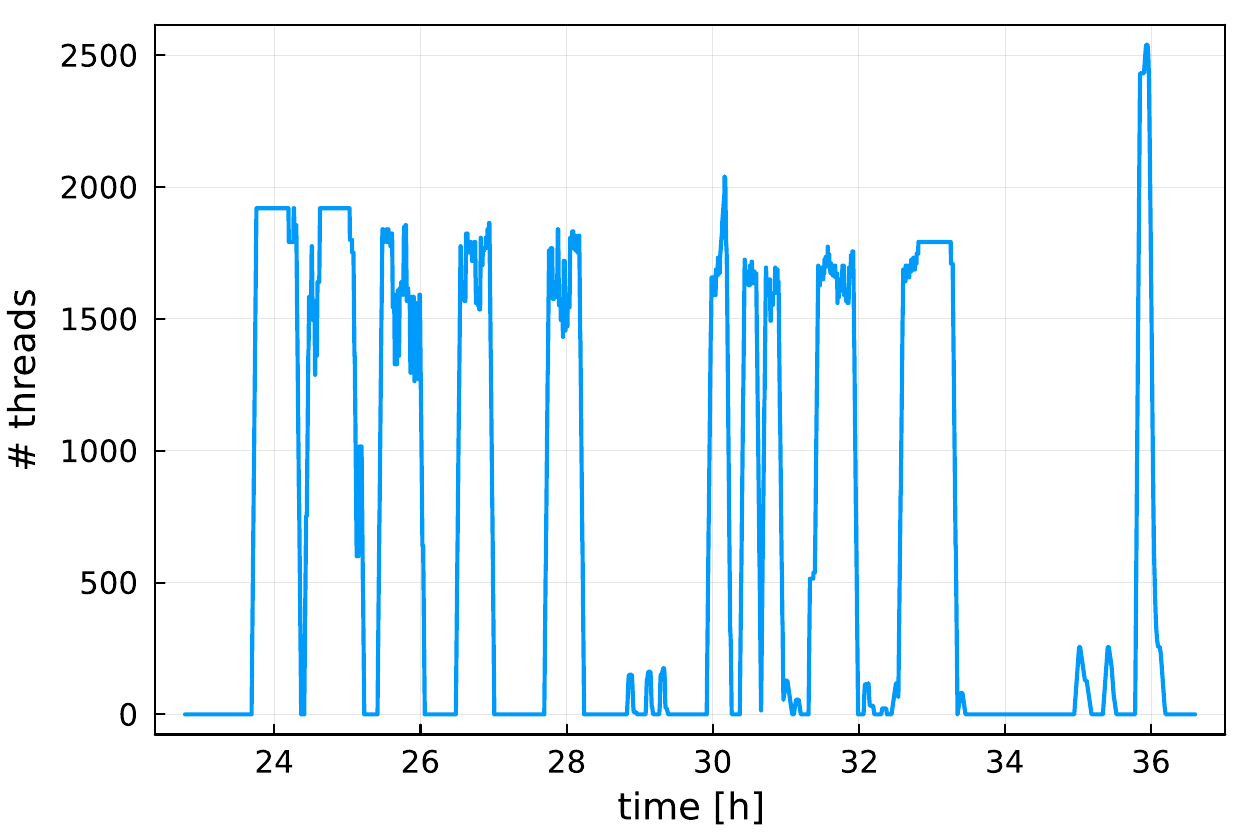}
    \caption{%
        CPU usage for the P175 experiment. \emph{Left:} CPU usage on Cori
        Haswell for the whole duration of the experiment. Only the day shifts
        collected data, therefore no data analysis was needed at night.
        \emph{Right:} CPU usage for one day shift (on the second day of the
        experiment). We see the ``bursty'' CPU utilization that results from
        urgent computing: whenever new data are available they need to be
        analyzed as quickly as possible. Once data have been analyzed, the CPUs
        on Cori go idle, while waiting for new data.%
        \label{fig:reservation}%
    }
\end{figure*}

Typical XFEL experiments involve collecting multiple datasets for the same or
similar samples, potentially moving them through some reaction condition and
capturing the structural changes as the reaction proceeds, or screening
proteins with a variety of ligands that are biologically or pharmacologically
relevant, such as in the case for the COVID-19 viral proteins from experiment
P175.  Samples therefore accrue a great deal of metadata, and each run needs to
be associated with this metadata so datasets can be produced from the right
subsets of diffraction images.  Therefore, the first task the user completes in
the \emph{cctbx.xfel} GUI is tagging runs with short, descriptive terms, such
as ``batch1'', ``reactionstate2'', or ``ligand3''. Multiple tags can be added
to a data set.

Next, the user needs to provide processing parameters for each dataset.  These
parameters include details needed to extract reflection data, the experimental
geometry such as the location of the detector in 3D space, and known crystal
properties.  These parameters will need to be updated (with better estimates)
as the experiment progresses, and so they are organized by trials, in which the
user can change the parameters and re-process the data. This organization into
trials is particularly helpful when keeping track of which parameters were used
during re-processing.

Finally, the user specifies which tags will form a dataset, mixing and matching
them as needed.  With these properties in place, the GUI will run through a
cycle of determining which tasks are needed to be performed on which data, and
submitting these tasks to the cluster to be processed.  The GUI monitors the
state of each task and continues to submit new jobs as data arrive or as
processing tasks finish, allowing downstream tasks to be submitted on upstream
results.

The \emph{cctbx.xfel} GUI therefore provides the experiment's operators with a
complete pipeline management tool, which lets multiple users simultaneously
specify analysis parameters and view analysis results. When new data or
analysis parameters are detected \emph{cctbx.xfel} automatically builds Slurm
job scripts and input files, and submits these to a set of reserved compute
nodes. Please see the video\footnote{%
  Video available  here:\,
  \url{https://doi.org/10.5281/zenodo.7439774}%
}
for a run-through of the \emph{cctbx.xfel} GUI\@. By acting as the interface
with the supercomputer, the \emph{cctbx.xfel} pipeline management system allows
scientists to treat HPC as a reactive element.  Fig.~\ref{fig:reservation}
shows a time series of the CPU utilization during the P175 experiment. This
usage pattern is typical of the \emph{cctbx.xfel} workflow: whenever new data
are available, they need to be analyzed as quickly as possible resulting in a
sudden need for up to 64 Cori Haswell nodes.

\subsection{Processing Data on Cori Compute Nodes}
\label{sec:data-analysis}

\begin{figure*}
    \centering
    \includegraphics[width=0.75\textwidth]{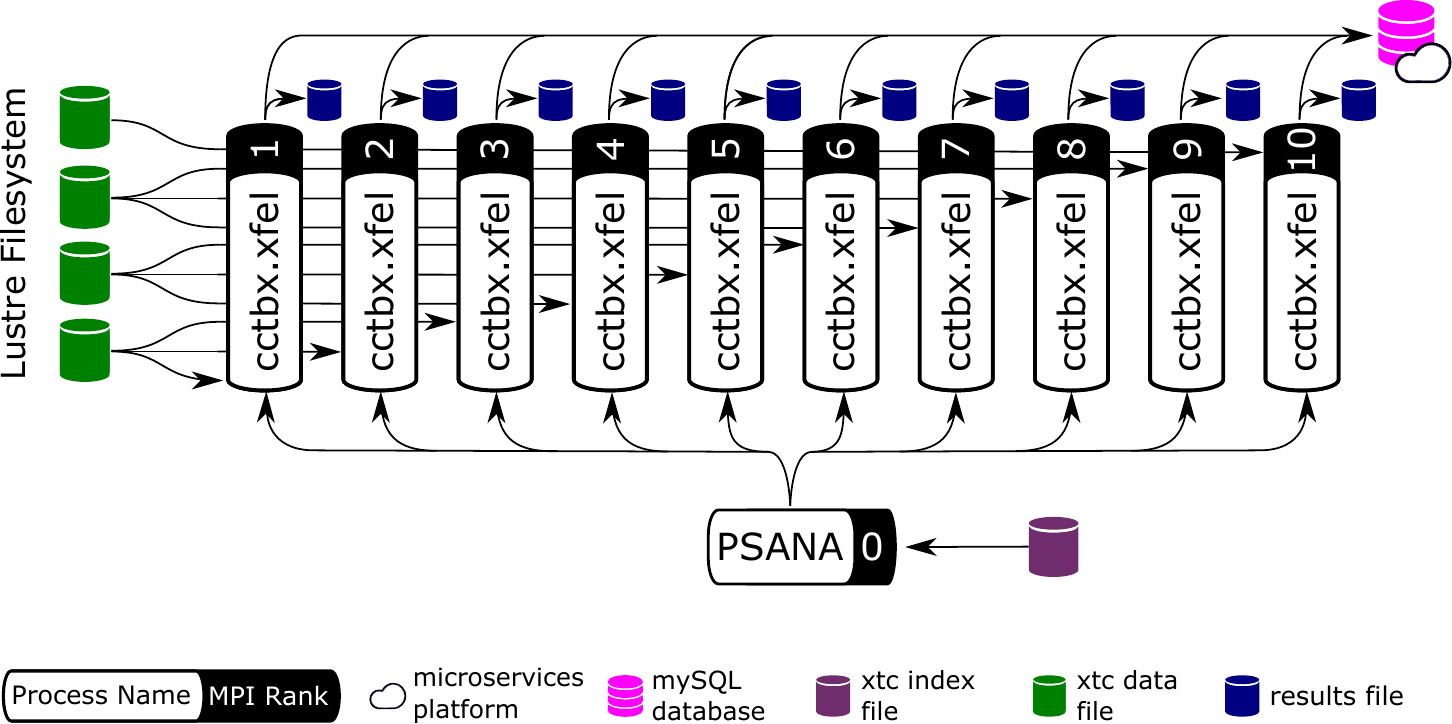}
    \caption{%
        Structure of an analysis worker running on the Cori Haswell nodes.  We
        rely on MPI parallelism to distribute work between nodes (OpenMP is
        also available, but was not needed to achieve the desired throughput).
        We employ a producer/consumer model to distribute work and achieve load
        balancing.  Data is provided by \emph{psana}, which runs on the first
        MPI rank. \emph{psana} reads an index file and distributes work to the
        \emph{cctbx.xfel} workers. The resulting program is a flat tree of MPI
        ranks with data analysis ranks located at leaves. Workers access data
        directly by reading the raw data files using offsets provided by the
        ``PSANA'' (root) tree node.  Finally, the \emph{cctbx.xfel} workers
        save their results to disk (local to each MPI rank, using the DataWarp
        burst buffer) and report the analysis progress to a MySQL database
        hosted on NERSC's Spin micro-services platform.  Arrows indicate the
        overall flow of data.%
        \label{fig:analysis_workflow}%
    }
\end{figure*}

Data were processed on up to 64 Haswell nodes  on NERSC's Cori XC40 system. The
computational workload is highly variable (\emph{cf}.\
Fig.~\ref{fig:reservation}) depending on the nature of the data being
collected. XFEL data analysis follows several sequential stages: i) Identifying
Bragg spots in a diffraction image (\emph{spot finding}); ii) Associating each
Bragg spot with a Miller index (\emph{indexing}); iii) Refining unknown model
parameters (\emph{refinement});  iv) integrating the Bragg spot intensities and
subtracting background (\emph{integrating}). and v) scaling each image and
combining measurements of the same Miller indices collected over several images
(merging). Table~\ref{tab:processing} shows that each stage is lossy: not every
image contains data of sufficient quality (\emph{ie}.\ not enough
high-intensity Bragg spots) to conclusively analyze. This means that each
subsequent stage processes fewer data -- and therefore needs fewer
computational resources. Hence stage (v) requries much smaller jobs on Cori
than stages (i)--(iv).

The computational motif is identical (except for number of images) for each
stage. \emph{cctbx.xfel} uses MPI to distribute work over up to 64 Cori Haswell
nodes. The work is distributed using a producer/consumer model, where each
image is processed largely independently. Fig.~\ref{fig:analysis_workflow}
sketches this computational motif. We use \emph{psana} \cite{psana2016} to read
the raw data files. In Fig.~\ref{fig:analysis_workflow} we show an example
configuration where rank 0 distributes work to available ranks. The producer
rank uses MPI to distribute offsets into the raw data files (green ``buckets''
in Fig.~\ref{fig:analysis_workflow}). The worker ranks then process each image
independently, by accessing the data files (each run's data is stored across
several files) using an offset and applying the detector calibration in memory.
From here on stages (i)~--~(iv) are applied without communicating with any
other ranks. Finally results are stored to the DataWarp burst buffer (blue
buckets) and a MySQL database (pink database icon in
Fig.~\ref{fig:analysis_workflow}).

\begin{SCfigure}[0.7]
    \centering
    \includegraphics[width=0.55\textwidth]{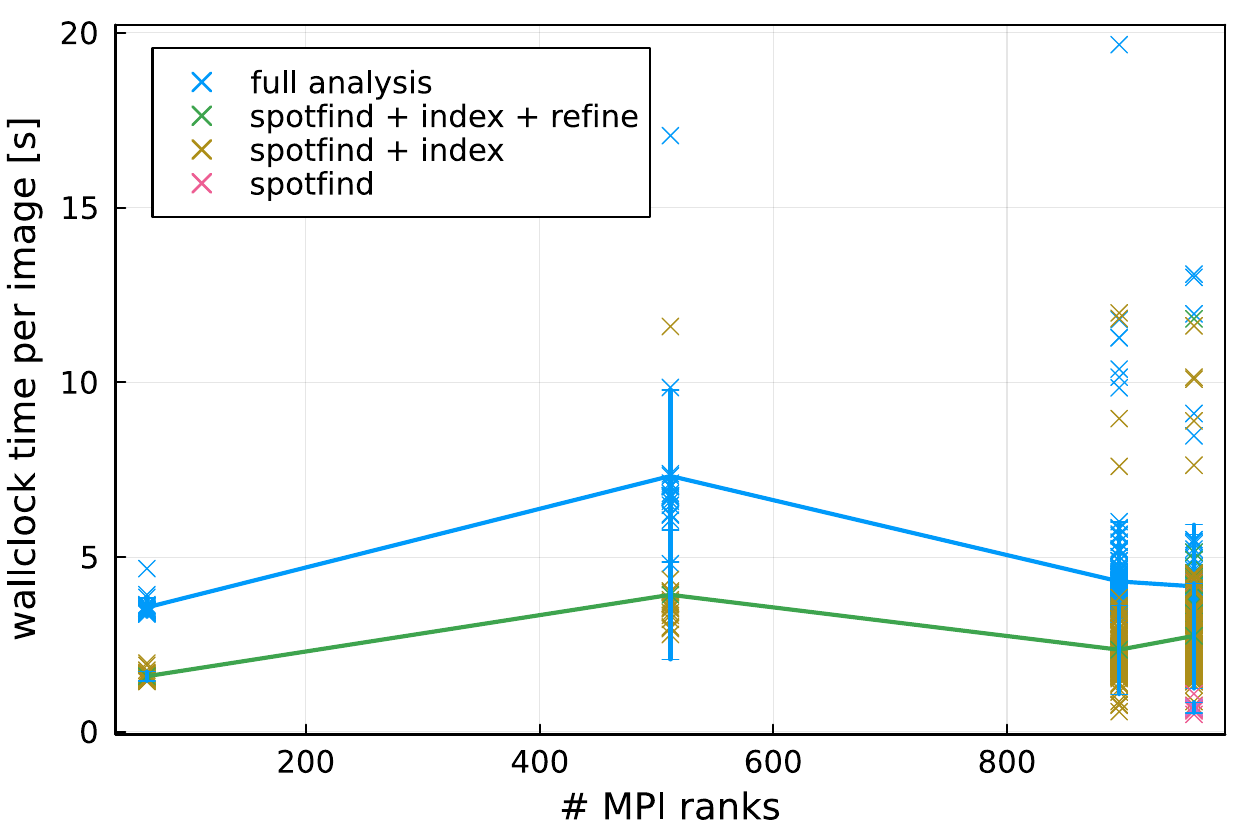}
    \caption{%
        The average time to process an image remains constant with the number
        of MPI ranks used. Colors show the different stages of the data
        analysis pipeline. We also see that the variability grows with number
        of MPI ranks, in part due to increased resource contention.  However,
        the vast majority of images can be processed with near-constant time,
        achieving weak scaling on the Cori Haswell nodes.%
        \label{fig:p175_scaling}%
    }
\end{SCfigure}

\begin{SCfigure}[0.7]
\centering{}
\includegraphics[width=0.55\textwidth]{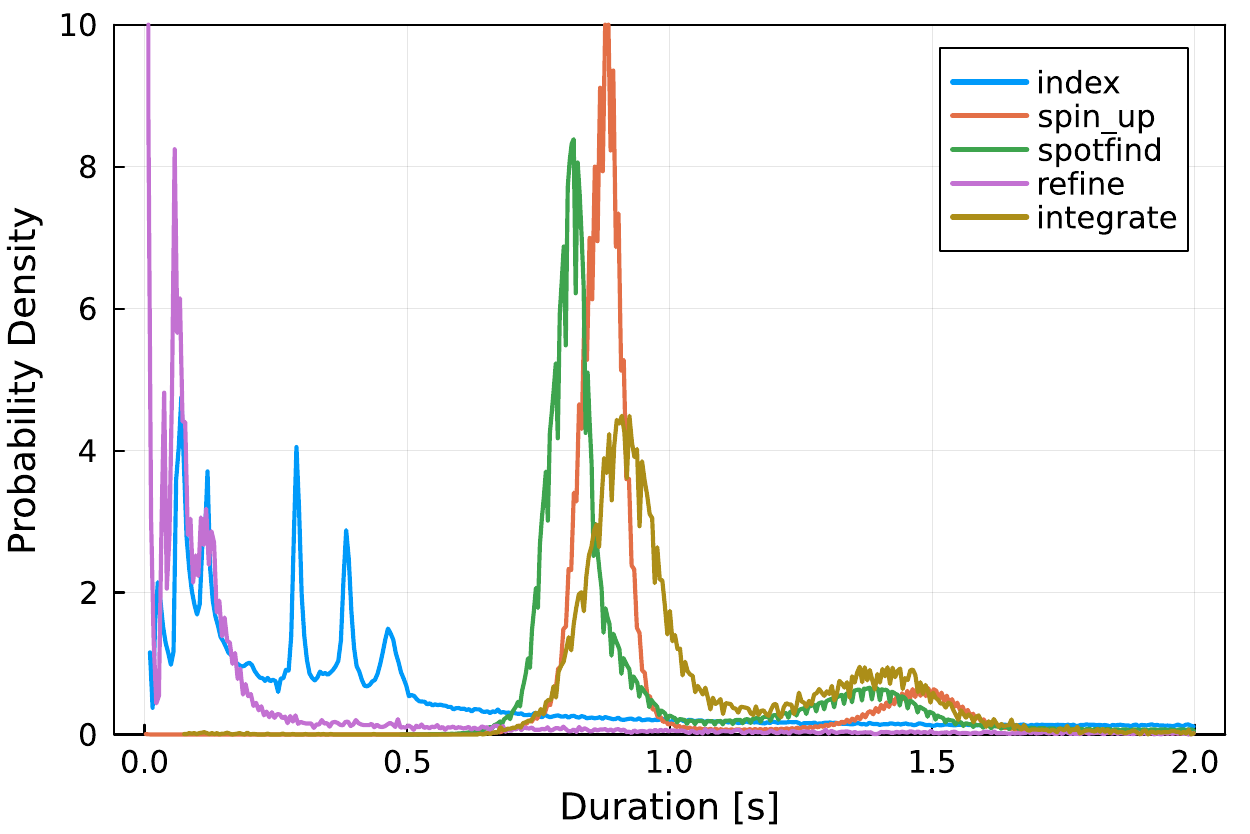}
\caption{%
    Probability distribution of the time taken to perform different data
    analysis tasks. While most processing steps complete within a few seconds,
    data analysis can occasionally take significantly longer. Due to this
    variability, our workflow uses producer/consumer
    parallelism~--~which is automatically load-balanced. In LV95, a fast
    algorithm was used for small molecule data\cite{schrieber2022}. For protein
    data, such as P175, it can take longer to index.%
    \label{fig:performance}%
}
\end{SCfigure}

Fig.~\ref{fig:p175_scaling} shows the average time to process an image. We see
that \emph{cctbx.xfel} achieves near-ideal weak scaling, regardless of whether
a partial data analysis (red, and green symbols), or a complete data analysis
is being performed. The performance variability in the wallclock per image does
increase with the number of MPI ranks. This is primarily due to shared resource
contention such as I/O and network latency. Fig.~\ref{fig:performance} shows
the probability density function of the wallclock time for each step. This
variability can have two sources: 1) algorithmic: \emph{e.g.}\ the peaks in the
green curve show different indexing algorithms being applied to the data (note
that to analyze LV95, a fast algorithm was used for small molecule
data\cite{schrieber2022}. For protein data, such as P175, indexing can take
longer.); and 2) resource contention. The distributions are strongly-peaked,
and therefore the vast majority of images are analyzed within 7s. However as it
is not possible to predict exactly how long it will take to analyze a batch of
images, we use a producer/consumer workflow as it is automatically
load-balancing.

\begin{figure*}
    \centering
    \includegraphics[width=1.00\textwidth]{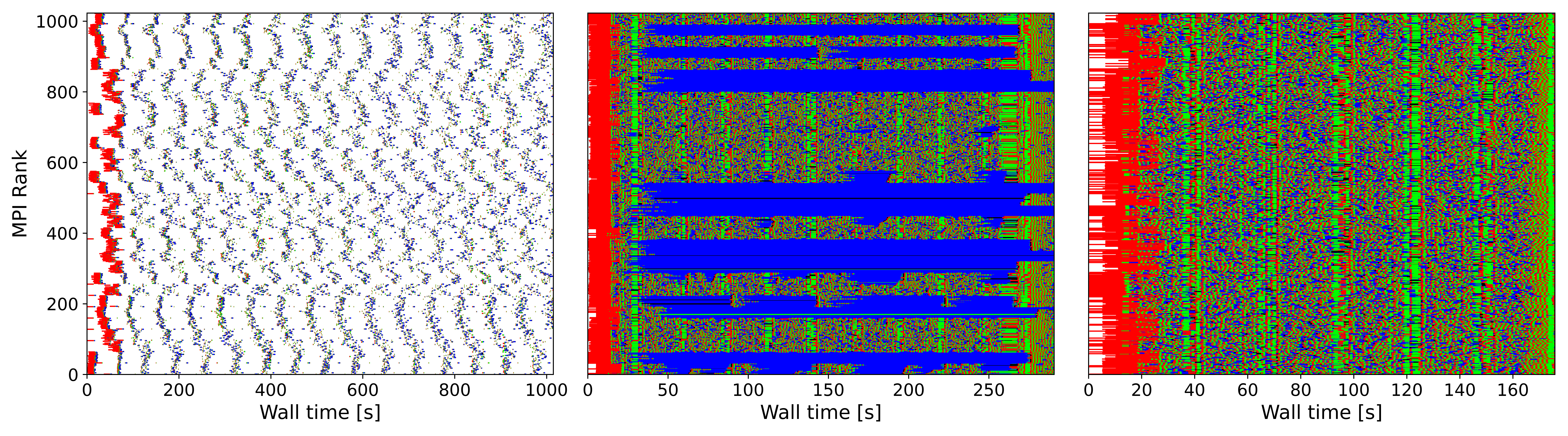}
    \caption{%
        Computational weather plot illustrating two barriers to scaling (left
        and center), as well as near-optimal performance (right). Weather plots
        show resource contention by plotting the data processing timeline of
        each rank. The colors represent different processing steps:
        initialization and I/O (red); spot finding (green); indexing (blue);
        model refinement (dark green); and integration (black). MPI
        communication is not profiled and is included in the white areas. The
        left plot shows an MPI communication-bound setup. Performance profiling
        revealed a load-imbalance, causing ranks to wait for MPI communication.
        After optimizing the MPI work sharing code almost all white space
        disappears. However this reveals I/O contention (note that all steps
        open files when saving intermediate results and for logging) on the
        \emph{SCRATCH} file system (central plot) as shown by some nodes
        working normally, while others appear stuck. Switching to the DataWarp
        burst-buffer resolves this I/O contention resulting in near optimal
        performance (right plot).%
        \label{fig:weatherplot}%
    }
\end{figure*}

A helpful tool to identify performance variability due to resource contention
is the computational weather plot as shown in Fig.~\ref{fig:weatherplot}. The
MPI ranks are enumerated on the $y$-axis and wallclock time is plotted on the
$x$-axis. Each worker is plotted as a collection of horizontal lines (a new
line for each image). As different images are analyzed, the horizontal line is
given different colors: initialization and I/O (red); spot finding (green);
indexing (blue); model refinement (dark green); and integration (black).  MPI
communication happens only when images are assigned to a particular worker and
therefore those regions are not plotted (\emph{ie}. they are the white regions
between images). Results are stored at the end of each processing step (raw
data files are only read during the initialization step).

To demonstrate this powerful diagnostic tool, the left and central panels shown
in Fig.~\ref{fig:weatherplot} show two different forms of contention. The left
panel shows an MPI communication-bound job: most time was spent between images,
waiting for new work (which is distributed using MPI). Performance profiling
revealed a load-imbalance, causing ranks to wait for MPI communication. The
central panel shows an example of I/O contention: at the end of each processing
step data is written to the Lustre \emph{SCRATCH} file system, which resulted
in several nodes hanging while trying to open files simultaneously. The right
panel shows the same setup where each rank caches results to the DataWarp burst
buffer instead of \emph{SCRATCH}. Note that all data processing steps open
files when saving intermediate results and for logging. We find that the use of
the burst buffer reduces this I/O contention. For a 32-node (1024 rank) job,
using the burst buffer therefore leads to a $1.8 \times$ performance speedup on
average.

\subsection{Workflow Orchestration}
\label{sec:workflow-orchestration}

\begin{SCfigure}[0.7]
\centering{}
\includegraphics[width=0.55\textwidth]{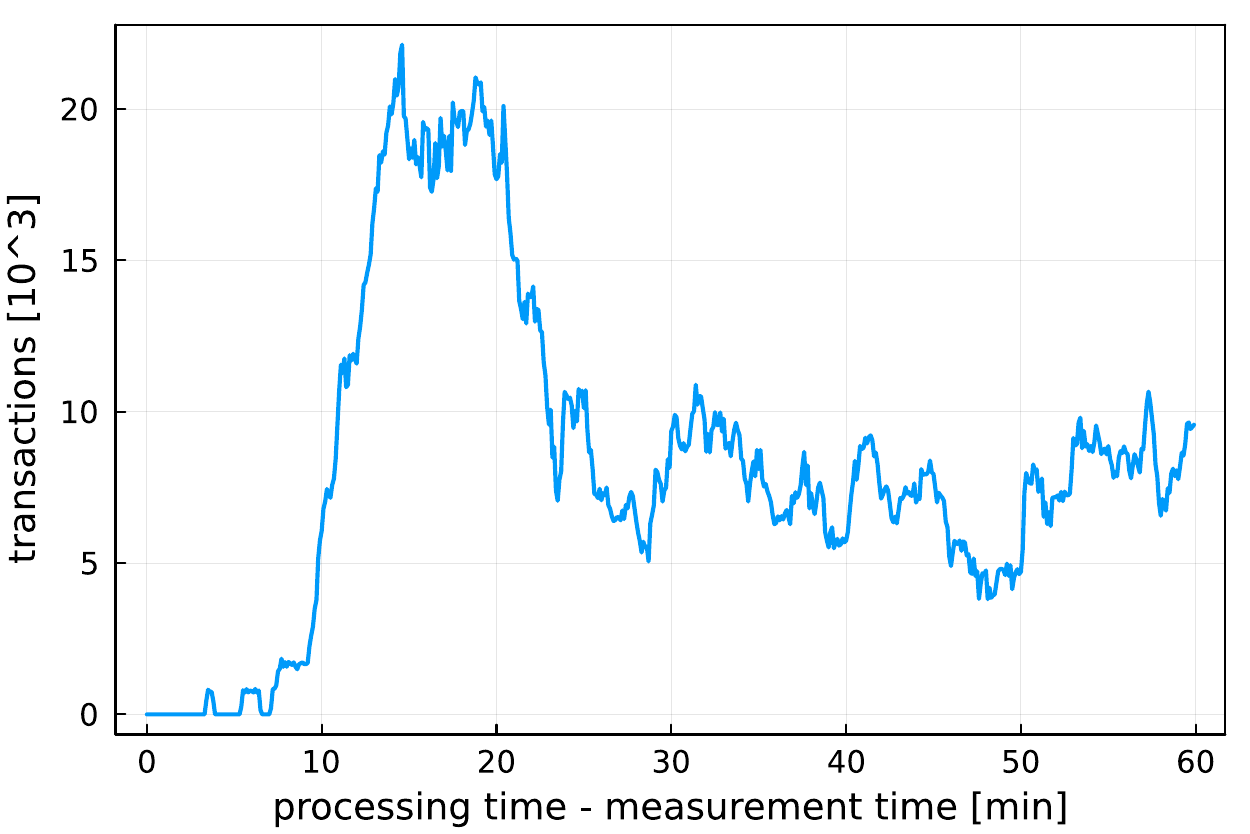}
\caption{%
    Delay time between recording an event and completion of the first data
    processing step for the P175 experiment. The graph shows the number of
    processed events (transactions), as a function of delay time. We find that
    a few images (those at the end of a run) are processed within 3.5
    minutes~--~given that a data transfer usually takes approx 3 minutes, these
    images were processed only a few seconds after arriving at NERSC\@. Most
    images were processed within approx. 10-20 minutes. Tailing delay times
    greater than 20 min are due to data reprocessing (\emph{cf.}
    section~\ref{sec:urgent-processing}).%
  \label{fig:transaction_pdf}%
}
\end{SCfigure}

We use a relational database implemented via MySQL to maintain the associations
between data and results in various stages of processing. In addition, since
processing results are logged to the database quickly, we can access those
results from the experiment control room and display them to on-site users for
rapid feedback on the data they are collecting.

We selected Spin, a NERSC microservices platform for container-based services,
to host our MySQL server reliably and scalably\footnote{%
    The scripts to deploy a MySQL database using Spin are available here:\,
    \url{https://gitlab.com/NERSC/lcls-software/-/tree/beamtime-2020-09/spin/mysql-p175}%
}.%
Spin hosts services that can be accessed from the Cori compute and login nodes.
Having access to both kinds of node is essential because the \emph{cctbx.xfel}
GUI runs on the login nodes and needs to be able to query the database in order
to display the progress of data processing jobs, as well as determine which new
jobs to submit. The workers do not query the database, instead they commit the
status of the images they are processing (\emph{e.g.}\ number of spots found
per image, the rate at which they are indexed, etc). Hence, even though
thousands of ranks will be committing status updates to the MySQL database,
these transactions are light weight, with the MySQL service handling them well.
We found that database connections and transactions consumed between 1\% and
3\% of total runtime. This includes latencies caused by accessing Spin via the
(slower) TCP network. Furthermore, Spin is scalable, which enables us to
flexibly increase the number of connections the database service can
efficiently manage as we scale to ever larger data processing workloads.

Fig.~\ref{fig:transaction_pdf} shows the effectiveness of this approach by
plotting a histogram of the difference between the data-collection time and the
processing time. We see that some images (a few thousand) were processed within
3.5 minutes -- this includes the transfer time of approx. 3 minutes. The
majority of images are processed between 10-20 minutes after data is collected.
While this does not include reprocessing, or interpreting the results, it does
demonstrate that cross-site automation is crucial for fast turn-around.

\section{HPC Challenges}

While this is a relatively modest computing footprint compared to traditional
HPC workloads, real-time data analysis requires the coordination of many moving
parts ranging from traditional computing to networking and I/O\@. Data sets are
expected to grow at least 3000-fold with increasing detector resolution, beam
intensity, and measurement rate\cite{superfacility,bard2022,antypas2021}.
Therefore, the performance profiling data we collected represents an important
benchmark, allowing us to extrapolate the overall performance of this
Superfacility workflow and predict future bottlenecks which would prevent
scaling.

\subsection{Urgent and Real-time Computing}
\label{sec:urgent-processing}

XFEL data reduction challenges computing clusters in two ways 1) unequal data
processing needs per frame and 2) stochastic (\emph{cf.}
Fig.~\ref{fig:performance}) and bursty (\emph{cf.} Fig.~\ref{fig:reservation})
computational needs. Together these result in a demand on the job scheduler
where computational resources are urgently needed (the urgency is due to the
need for fast real-time data processing), with little advance warning (only
after all the data has been processed, do we know how many images resulted from
``good'' measurements).

At NERSC we have enabled time-sensitive computing by allowing nodes on Cori to
be \emph{reserved} ahead of time. These nodes will then be kept clear of jobs
not explicitly submitted to this reservation.

\subsubsection{Unequal Data Processing Time}

\begin{figure}
    \centering
    \includegraphics[width=0.75\textwidth]{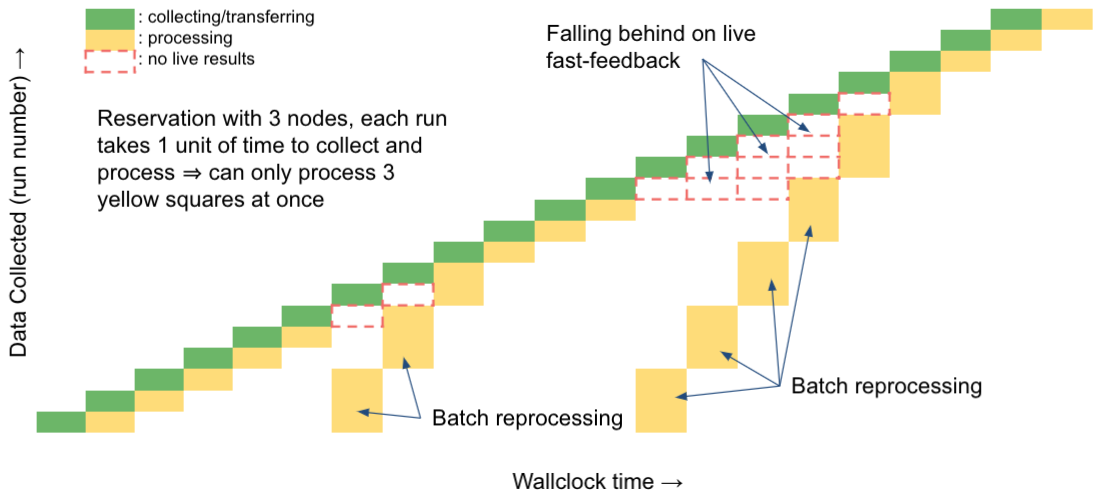}
    \caption{%
        An illustration of the XFEL urgent computing needs. The $x$-axis
        represents time, and $y$-axis represents the number of data sets
        collected. To keep up with processing as data from runs arrive (green
        boxes), processing jobs are submitted as soon as possible (yellow
        boxes). For simplicity we assume that it takes roughly the same amount
        of time to process a data set as it takes to collect it (green and
        yellow boxes are the same size). Furthermore, to illustrate the problem
        of limited reservation sizes, we assume that our reservation has a
        maximum size of 3 nodes (3 yellow boxes). When new parameters are
        discovered, all data must be re-processed in a batch, and on a limited
        reservation, this can lead to delays in live feedback (red boxes).
        Furthermore, the burden of reprocessing grows with the data set size.
        Therefore a reservation would potentially need to be as large as the
        final data set.%
        \label{fig:batchreprocessing}%
    }
\end{figure}

In each run, thousands of image frames are recorded, but how far each frame
makes it through the processing pipeline varies widely.  A frame could be a
complete miss, without a crystal.  A crystal may not be of sufficient quality
to be processed, and even if it is, it may not be isomorphous with the rest of
the data.  At each step, the image can be rejected for a variety of reasons.
This is illustrated in Table~\ref{tab:processing}: each processing stage (row)
has a finite ``success'' rate, and therefore only a fraction of images go onto
the next stage.

As described in section \ref{sec:data-analysis}, we solve this problem by
splitting the pipeline into tasks and using fewer cores for downstream tasks.
For example, during P175, for indexing and integration, we used 28 nodes per
job, but for scaling and merging which does not read the pixel data, we only
used 1-2 nodes per job.

Further, for indexing and integration we use a producer/consumer approach,
where a root MPI rank sends images to the other ranks.  Each rank reports back
when they finish an image and receive a new one to process.  In this way, all
the ranks are kept busy until the images have all been processed.

\subsubsection{Stochastic and Bursty Compute}

Ideal processing parameters are rarely known when an experiment begins, and
midway through data collection, new parameters can be discovered which obviate
all previous processing results. In classical computing scheduling this leads
to two inefficiencies (Fig.~\ref{fig:batchreprocessing}). First, if the set of
reserved compute nodes is big enough to accommodate processing needs plus an
additional safety margin, then when data is not being collected or when typical
processing patterns are being observed, the cluster can be underutilized.
Second, when batch-reprocessing needs to occur due to the addition of new
parameters, real-time processing can fall behind. 

These problems necessitate different scheduling systems than reservations or
first in-first out. Our experiences with real-time data processing for LV95 and
P175 have shown that reservations are able to guarantee enough computational
resources for time-sensitive data processing. However reservations alone can be
a wasteful solution: any time the reservation goes unused (\emph{e.g.} between
measurements) will result in idle compute nodes.

A more efficient arrangement could include a mix of reservations plus real-time
priority access to compute resources, which can be released to lower-priority
jobs when not being immediately used. Furthermore, preemption is a promising
solution to allow underutilized reservations to be filled by preemptible jobs
until the compute nodes are needed for urgent computing tasks. Preemptible jobs
are programs that listen for a system interrupt (\emph{e.g.} SIGINT),
and~--~upon receipt~--~gracefully save and quit. At NERSC, together with
SchedMD, we have developed a reservation system by which preemptible jobs can
enter a reservation\cite{bard2022,giannakou2021}. These will then be stopped if
new jobs are submitted directly to the reservation (after a warning period
during which SIGINT is used to request that the job saves and quits). This
technology is in its early stages, and we describe our initial experiences with
preemptible reservations in \cite{giannakou2021}.

\subsection{I/O and Network Performance}

\begin{SCfigure}[0.7]
    \centering
    \includegraphics[width=0.55\textwidth]{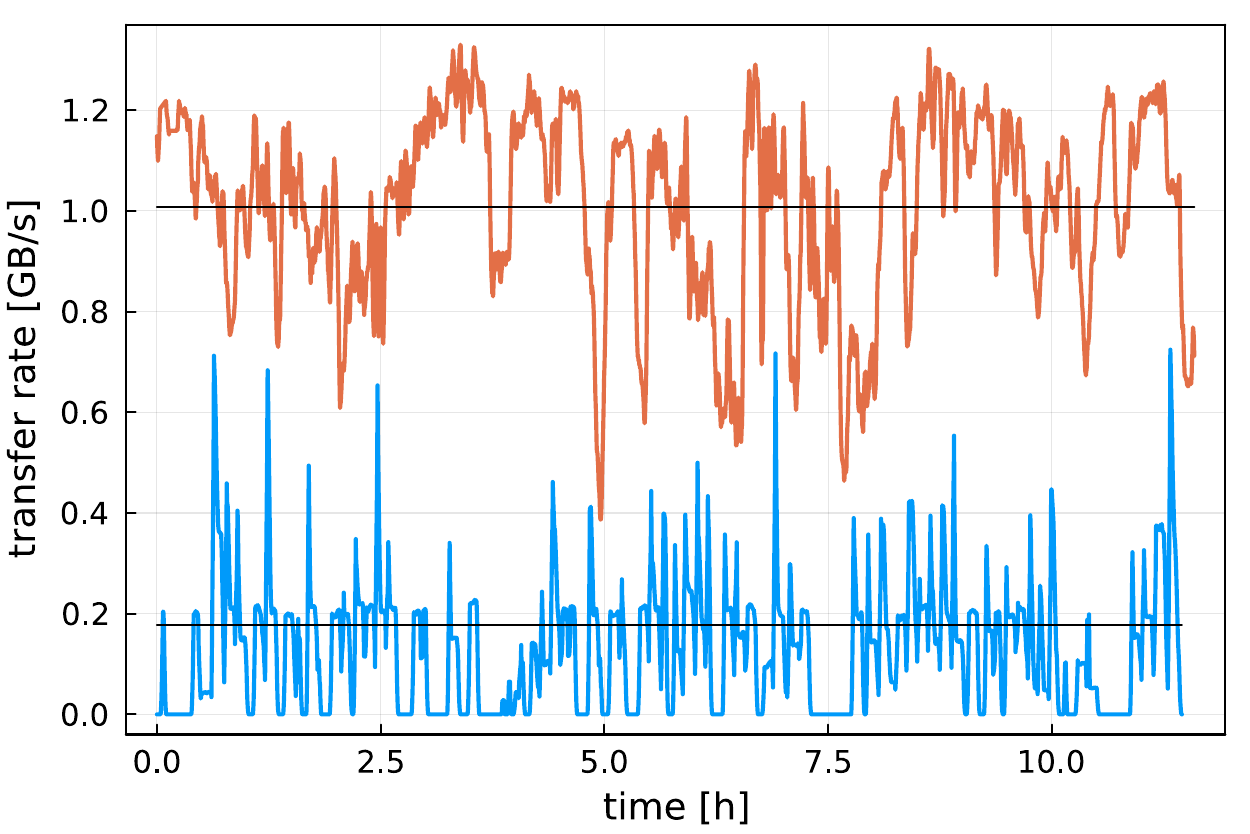}
    \caption{%
        Data transfer rate between the LCLS and NERSC (Lustre \emph{SCRATCH})
        file systems. This rate includes disk read and write speeds, which
        ultimately limited the rate at which data can be transferred to
        NERSC\@.  Horizontal black lines show average transfer rates. The
        orange line shows a representative ``good'' data transfer speed.
        However, depending on contention in the Lustre file system at NERSC,
        this transfer rate can be $5-6\times$ lower -- shown by the blue line.%
        \label{fig:xrootd_comparison}%
    }
\end{SCfigure}

The 100 Gb/s network connection (hosted by ESnet) between LCLS and NERSC made
it possible to transfer most raw data files within 3 minutes after concluding
the run. Bandwidth on the ESNet link was reserved ahead of time using the SENSE
API\cite{sense}.  The XRootD clusters at LCLS and at NERSC performed well, and
could be scaled easily to accommodate more files if a backlog occurred.

In fact, the I/O speeds of the Lustre file systems at LCLS and at NERSC were
the rate limiting factor. Fig.~\ref{fig:xrootd_comparison} shows the end-to-end
data transfer rate from LCLS to NERSC\@. The different lines are measurements
taken on two different days. This makes it clear that there are ``good'' and
``bad'' days for file system utilization.  The 5-6 fold difference is due to a
bug in NERSC's \emph{SCRATCH} file system, where some of Lustre's Object
Storage Targets (OST) have a slow write speed. On a ``bad day'' the slow Lustre
write speed can become the dominant bottleneck in the data processing pipeline,
where the data for run $N$ is not transferred before run $N+1$ commences. This
highlights that reliable high-performance I/O is crucial for experimental
science workflows.

\subsection{Workflow Orchestration}

\begin{SCfigure}[0.7]
    \centering
    \includegraphics[width=0.55\textwidth]{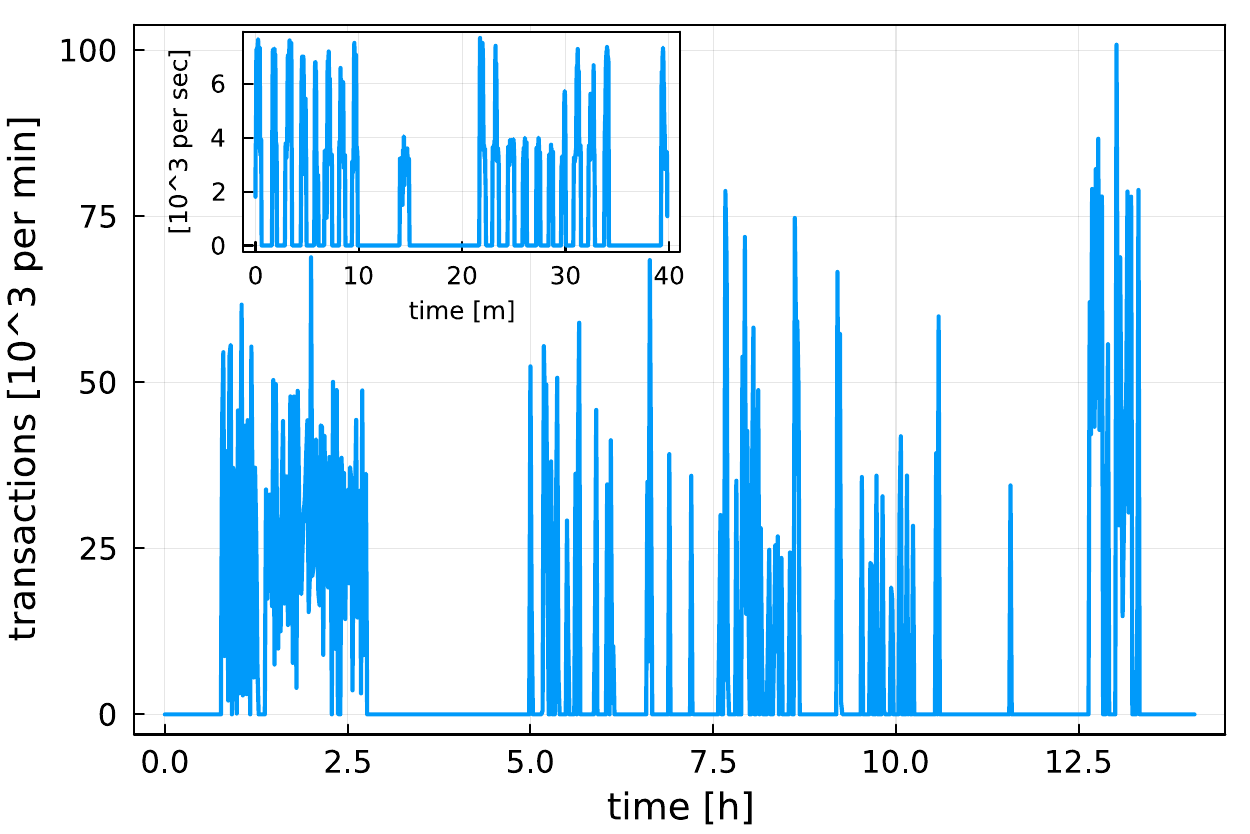}
    \caption{%
        Rate of database transactions during live data processing. The main
        plot shows the number (in thousands) of database transactions per
        minute during a 12 hour shift. The inset shows a 40-min snapshot of
        number (in thousands) per second. We see that the database receives up
        to 8000 commits/second, whenever data processing takes place (the
        ``bursts'' in the inset show individual data analysis jobs). Despite
        this heavy load, the Spin microservices platform was capable of
        handling this load level.%
        \label{fig:db_transactions}%
    }
\end{SCfigure}

Workflow orchestration at scale is always a challenge, as potentially hundreds
of thousands of tasks need to be coordinated from a central place. In our
workflow manager, the database takes on the role orchestrating the distributed
data processing. Therefore database communication is a potential single point
of failure and a bottleneck when experiments are scaled up to the kHz regime
with thousands of MPI ranks reporting results simultaneously.
Fig.~\ref{fig:db_transactions} shows that a Spin-hosted MySQL database was able
to accommodate the load of approx.  8000 transactions/sec.

While the MySQL database server was selected with scaling in mind, some further
optimizations became necessary when performing large-scale analysis runs.
\begin{enumerate}
    \item Limiting concurrent connections: In some configurations, the usable
        number of MPI ranks was limited by the concurrent connections that our
        database could support. We refactored our database communication to
        cache all database queries for a small set of images before flushing
        the cache via a single temporary database connection. This reduced the
        peak concurrent database connections to $~$1 per 10 MPI ranks.
    \item Transactions: For processing in the kHz regime during a different
        experiment we encountered another bottleneck when many small queries
        had to be executed sequentially, with later queries depending on
        earlier ones. Without access to the Spin system, we were overloading
        the MySQL server , to the point where logging 50000 images could take
        over an hour.  Using the MySQL statement LAST\_INSERT\_ID() we were
        able to combine many queries into a single transaction. With this
        approach, we could log these images using a single MySQL query
        comprising 130K lines that takes 0.07 seconds. 
\end{enumerate}

A related challenge to workflow orchestration is the variable processing time
per image. Fig.~\ref{fig:performance} shows the variability due to
\emph{algorithmic} differences between images (\emph{e.g.} the peaks in the
green line are due to the indexing algorithm ``trying'' different approaches to
find a solution). Therefore we employ a producer/consumer model to distribute
parallel tasks across MPI ranks while maintaining a balanced workload
(\emph{cf.} section~\ref{sec:data-analysis}). As the data analysis for each
image can have a subtly different call tree, this can have a subtle impact on
optimizing performance and diagnosing errors: we can not expect each logical
task to take roughly the same amount of time. We observe that between 2\% and
3\% of images take significantly longer than 2s to process. Using the {\it
hatchet}\ tool \cite{hatchet} we were able to compare the profiles for jobs
with different call trees. Hatchets allows us to analyze each job's call tree
hierarchically, and compare common sub-graphs. We found that the slow jobs were
a result of I/O contention while reading data, saving results and logging
progress. This highlights an important difference to many simulation codes:
data analysis workflows often have branching source codes, and invoke many
libraries~--~it is therefore not always possible to optimize the overall run
time by merely focusing on a handful of subroutines that are called over and
over.

\section{Superfacility API}

Over the years, NERSC staff have observed how many research workflow operations
fall into natural patterns of recurring actions that are carried out when
analyzing data. The traditional approach for HPC centers is to provide
human-readable interfaces and also to design the experience to meet the
interactive expectations of a human user. However this design collapses with
workflows that need to run at larger scale or at faster rates such as
automated, machine-driven workflows initiated at external facilities such as
LCLS-II\@. We expect this mode of operation to become more prevalent in the
future as more and more DOE facilities intend to link into ASCR computing
infrastructure to address their data and computing needs. Providing
machine-readable APIs for HPC resources is the logical prerequisite to make
this connection happen. It is also particularly fitting these days as the
workflows community comes  together to discuss common needs which, in turn, can
inform the development of such APIs
\cite{ferreira_da_silva_rafael_2021_4915801} .

Providing a modern API into NERSC is a central component of the Superfacility
project \cite{superfacility} at Lawrence Berkeley National Laboratory (LBNL),
which aims to lay the basis for a more unified, seamless environment that
combines hardware solutions, application software, and data management tools to
deliver breakthrough science. Automation is a key component of the
Superfacility concept, which envisions science teams at experiment facilities
orchestrating automated data analysis pipelines which move data from the
instrument to the computing site, perform analysis, and disseminate
results~--~all without any human in the loop.

The SF API provides RESTful API interfaces to resources and takes inspiration
from work at various HPC centers\cite{agave,firecrest} as well as from NERSC's
first API, the \emph{NE}RSC \emph{W}eb development \emph{T}oolkit
(NEWT)\cite{newt}. While NEWT was designed to serve primarily as backend
service for web science gateways, the new SF API is more targeted at workflows
and provides a modern, token-based authentication mechanisms as well as
asynchronous task execution.  The SF API service itself is built as a set of
Docker\footnote{\url{https://www.docker.com}} containers and runs in
Spin\footnote{\url{https://www.nersc.gov/systems/spin/}}, NERSC’s
Containers-as-a-Service platform.  By and large, it orchestrates connections to
backend systems and databases, asynchronously manages any long-running tasks,
handles authentication and authorization, and hosts its own documentation.
Currently, the API provides the endpoints described in
table~\ref{table_endpoints}. As the API is in active development, the most up
to date documentation can be obtained online at the automatically generated
Swagger page.\footnote{%
    Superfacility API documentation generated using the Swagger toolset,
    available at\, \url{https://api.nersc.gov/api/v1.2/}%
}

\begin{table}[!b]
\renewcommand{\arraystretch}{1.3}
\caption{API endpoints.}
\label{table_endpoints}
\centering
\begin{tabular}{m{0.1\textwidth}  m{0.3\textwidth}}
\hline
\textbf{\texttt{/meta}} & information the API installation at the HPC center\\
\textbf{\texttt{/account}} & retrieve allocation info for a user or project \\
\textbf{\texttt{/utilities}} & browse, upload, and download files or a free form command \\
\textbf{\texttt{/storage}} &  move data between sites with Globus, or between NERSC storage tiers \\
\textbf{\texttt{/status}} & retrieve system health status, including planned outages \\
\textbf{\texttt{/compute}} & submit and manage jobs, check job status \\
\textbf{\texttt{/tasks}} & information about pending and completed tasks \\
\hline
\end{tabular}
\end{table}

Enumerating all of the use cases for the API would be too much to cover in this
manuscript as NERSC envisions \emph{all} of the common interactions with its
systems to become automatable. Instead, we close with describing two use cases,
where one describes the abstract case of checking system health before a file
transfer and the other describes a current application of the SF API in the
AutoSFX pipeline of LCLS-II (a similar pipeline as {\it cctbx.xfel}\ for serial
femtosecond crystallography data analysis).

\subsection{Example: Checking system health before data transfer.}

Because the demand for compute capacity is driven by detector output that can
vary cyclically, experiments often need HPC-scale computing at short notice.
Some experiments may have even arranged for multiple compute sites to be
available to handle workloads in a given time period. To build a truly
automated and resilient workflow, scientists need to be able to query the
health and status of a facility and make decisions based on the response; for
example, if a file system is unavailable, the workflow pipeline should choose
not to send data to it.  To assess the status of a NERSC resource, the API
provides the {\textbf{\texttt{/status/}}} endpoint. Keeping with the example of
an imminent file transfer, the workflow could query
{\textbf{\texttt{/status/dtns}}} and
{\textbf{\texttt{/status/community\_filesystem}}} in order to find out the
health of NERSC's data transfer nodes and the community file system,
respectively. A json-formatted return of one of those queries would look like
this:
\begin{verbatim}
{"name": "dtns",
 "full_name": "Data Transfer Nodes",
 "description": "System is active",
 "system_type": "filesystem",
 "notes": [],
 "status": "active",
 "updated_at": "2021-05-21T07:55:00"}
\end{verbatim}
A status indicated as "active" would now inform the workflow that the resources
is operational and that it could start the data transfer. It could use its own
tools for these transfers, but the API also provides the
{\textbf{\texttt{/storage}}} endpoint to move data between
Globus-enabled\footnote{\url{https://globus.org}} sites and between the NERSC
storage tiers. For planning further ahead, a query to
{\textbf{\texttt{/status/outages/planned}}} would provide any scheduled outage
in the future and would enable the workflow manager to choose an alternative
destination or date for the transfer.

\subsection{Example: Using the SF API in the LCLS AutoSFX pipeline.}

The LCLS data management system invokes the SF API to integrate its automation
engine (ARP) with NERSC computing resources. Data management events (start/end
runs, file transfers etc) automatically trigger analysis jobs, which are then
initiated, monitored and managed at NERSC using
\textbf{\texttt{/compute/jobs/cori}} calls.  Runtime progress bar updates from
the jobs, in addition to job statuses from
{\textbf{\texttt{/compute/jobs/cori}}}, are then pushed to the browser and
dynamically update the web UI\@. The entire AutoSFX workflow, consisting of
multiple index/merge steps, is expressed as an AirFlow Directed Acyclic Graph
(DAG). Each node in the DAG is executed by the ARP by composing
\textbf{\texttt{/utilities}} and \textbf{\texttt{/compute/jobs/cori}} calls
(see table~\ref{table_endpoints}). Summary results (for example, electron
density maps) are copied back to the experiment folders using
\textbf{\texttt{/utilities/download}} calls and displayed in the web UI\@. As
many of these calls target asynchronous endpoints (\emph{e.g.}
\textbf{\texttt{/compute/jobs}}) where each POST call generates a task, the
workflow frequently queries the \textbf{\texttt{/tasks}} to inquire the status
of those tasks in order to advance in the DAG.

\section{Conclusion}

In this paper we have demonstrated the power and possibility of using on-demand
HPC to analyse data in real time for a running XFEL experiment at LCLS\@.  This
will provide a new mode of sustainable operations for high data-rate
experiments (over 400$\times$ the rate of today's experiments) expected to come
online in 2025.  To achieve on-demand and real-time feedback for experiment
control, we have addressed scaling problems in the application, work
scheduling, data management and workflow management.  We have identified areas
for future development based on a series of carefully-profiled experiments
performed in late 2020, which achieved the goal of having the analysis keep up
with the experiment operation.  Most importantly, the experiments described in
this paper were not one-off demonstrations, but the start of a regular mode of
joint operations between an experimental user facility and an HPC user facility
that is both sustainable and scalable. HPC centers are increasingly being used
for this kind of experiment-driven workflow, and the tools and techniques
developed in this work were designed to be generalizable to other science
areas.

\section{Acknowledgements}

N.K.S. acknowledges support from National Institutes of Health grant GM117126.
N.K.S, J.P.B.,  and D.B. acknowledge support from the Exascale Computing
Project (grant 17-SC-20-SC), a collaborative effort of the Department of Energy
(DOE) Office of Science and the National Nuclear Security Administration. Data
were collected at the Linac Coherent Light Source (LCLS) at the SLAC National
Accelerator Laboratory, supported by the DOE Office of Science, OBES (contract
No. DE-AC02-76SF00515), and processed at the National Energy Research
Scientific Computing Center, supported by the DOE Office of Science under DOE
contract DEAC02-05CH11231.

\bibliography{references}

\end{document}